# Magnetic properties of nitrogen doped diamond: A first principles study

Homnath luitel[1,+,*]

[1]*School of Physics, University of the Witwatersrand, Johannesburg, 1 Jan Smuts Avenue, Wits 2050, South Africa*

\* E-mail: luitelhomnath@gmail.com

**Abstract – The m**agnetic and electrical properties of nitrogen doped diamond system have been studied within the framework of a density functional theoretical approach. Spin-polarised calculations reveal that only the nitrogen doped system with adjacent carbon vacancies (NV-centre) leads to stable magnetism in N-doped diamond. The magnitude of the induced magnetic moment increases linearly with increasing number of NV centres in the system. This result explains earlier experimental reports suggesting unconventional magnetism in N-doped pristine diamond. The 2p-orbital electrons of the three carbon atoms adjacent to the vacancy contribute to the magnetic moment in the system. Notably, lone N at the lattice site in diamond system fails to induce any significant moment, whearas the C-vacancy and N-interstitial positions induce magnetic moment in the diamond system. Moreover, the NV-center system has p-type semiconducting characteristics, making it a potential candidate for spintronics applications. To quantify the feasibility of different systems, the magnetic moment, ground state free energy, Fermi energy and defect formation energy were calculated for each structure.

---

[+]Present Address: Department of Physics, Nar Bahadur Bhandari Government College, Tadong, Gangtok 737102, India

**Introduction** – Over the past few decades, diamond based research has particularly gained paced due to its potential applications in quantum computing and spintronics devices [1]. To name few, nitrogen-doped diamond has applications in nanoscale magnetic sensing, magnetic resonance, and wide-field magnetic microscopy [2, 3].The sensitivity of NV-based diamond magnetometers can be improved via high-permeability magnetic flux concentrators, leading to a sensitivity of ~0.9pTs1/2 to magnetic fields in the frequency range between 10 and 1000 Hz [4]. NV centers in diamond have potential applications in highly sensitive magnetic sensors and quantum devices [5].

Nitrogen-vacancy (NV) centers in diamond are created via methods such as low-energy nitrogen implantation, carbon irradiation of nitrogen-rich high-pressure high-temperature (HPHT) diamond, and low-energy N+ or CN- implantation into nitrogen-free chemical vapor deposition (CVD) diamond [6, 7]. These methods affect the magnetic sensitivities of the resulting NV layers, with low-energy irradiation of HPHT diamond being competitive for creating thin NV layers for wide-field magnetic imaging [3]. N-implanted nanocrystalline diamond samples have been reported to be ferromagnetic [8]. However, strong existence of superparamagnetism has been reported in nitrogen-doped single crystalline diamond samples at low temperatures, which get vanished at temperatures above 25 K [9]. Despite all the attempts to understand the NV centers, the magnetic properties of nitrogen doped diamond remain unsolved and various reports on this topic are rare. Thus a detailed first principles analysis is neded to understand the origin of the magnetic moment.

Density functional theory based calculations have been proven to correctly predict the magnetic properties of various materials [10-12], which are in good agreement with experimental observations [13-15]. We used the standard theoretical approach under the framework of density functional theory to simulate the magnetic properties of a nitrogen doped diamond system. Nitrogen doped diamond along with various possible configurations has been considered for a detailed understanding.

**Computational Methodology** – Theoretical calculations in the framework of density functional theory were carried out using Quantum Espresso package [16, 17]. A unit cell of diamond with cubic crystal symmetry (F3dm) containing eight atoms and lattice parameter a = b = c = 3.567 Å [18] was multiplied along the three crystallographic axes to build a $2 \times 2 \times 2$ supercell consisting of 64 atoms ($C_{64}$). A N-doped diamond structure ($NC_{63}$) was formed by replacing a carbon atom in the pristine supercell. Aditionally, a N-V center ($NC_{62}$) is formed in the diamond supercell by removing an adjacent atom close to the dopant. Similarly, diamond systems with carbon vacancies ($C_{63}$) and nitrogen at interstitial position ($NC_{64}$) have been formed as shown in Figure 1.

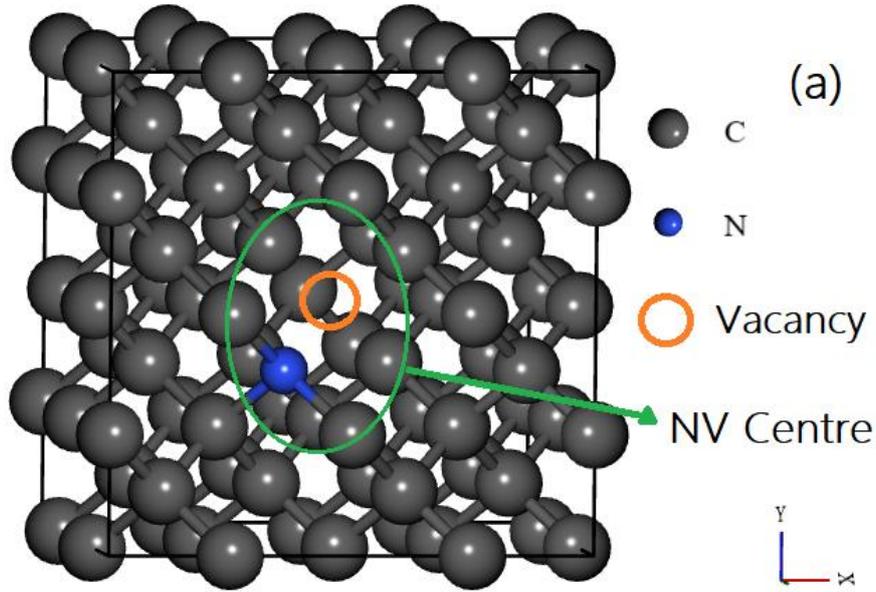

Figure 1: A 2x2x2 diamond super cell system with NV center ($NC_{62}$).

Throughout the calculation, Perdew–Burke–Ernzerhof revised for solids (PBEsol) is applied as the exchange–correlation functional along with the GGA approximation [19, 20] to study the spin polarised states of the pristine and other doped systems under consideration. The values of the Hellmann-Fynman force has been fixed to the limit of 0.001 Ry, the mesh cut off energy to expand the plane waves was allowed up to 65 Ry with the tolerance threshold limit of $10^{-7}$ Ry to stop the self-consistent field (SCF) calculations. During structure optimisation, only the atoms were allowed to relax in their lattice sites. The spin polarised scf calculations were carried out with a $6 \times 6 \times 6$ Mesh-Grid in the first Brillouin zone following Monkhorst-Pack (MP) method [21]. Similarly, a larger $3 \times 3 \times 2$ supercell consisting of 144 atoms has been considered for generating 2NV-center ($N_2C_{140}$) and 3NV-center ($N_3C_{138}$) systems. The larger supercell allows sufficient distance between two consecutive NV centers maintaining the periodicity condition in the DFT calculations. The total ground state energy was compared for the spin polarised state to find the stable ground state for all the structures.

**Results and Discussions** – Origin of magnetic property in nitrogen doped diamond system was determined under the framework of density functional theory. Atomic vacancies play a major role in the magnetic properties of different functional materials [22, 23]. Thus, diamond systems with different possible configurations viz, $C_{64}$, $NC_{64}$, $NC_{63}$, $NC_{62}$ and $C_{63}$ have been considered to incorporate the effects of atomic vacancies. The details of the magnetic moment, Fermi energy change, defect formation energy and ground state free energy for each diamond system under consideration are tabulated in Table 1. Interestingly, nitrogen when doped at the lattice site of diamond, does not induce any magnetic moment unlike other materials [24, 25]. However, the nitrogen doped system along with adjacent carbon vacancy (NV-center) generates a net magnetic moment of 1.12 $\mu_B$ in the system. The N-atom present in the lattice gives rise to a very small moment, where the major contribution comes from the three immediate neighboring carbon atoms surrounding the vacancy. Furthermore, the ground state free energy of magnetic state is approximately 63 meV less than that of the non-magnetic state, indicating stable magnetism in the NV-centerd diamond system. A comparison of the induced magnetic moment in various configurations is shown in Figure 2.

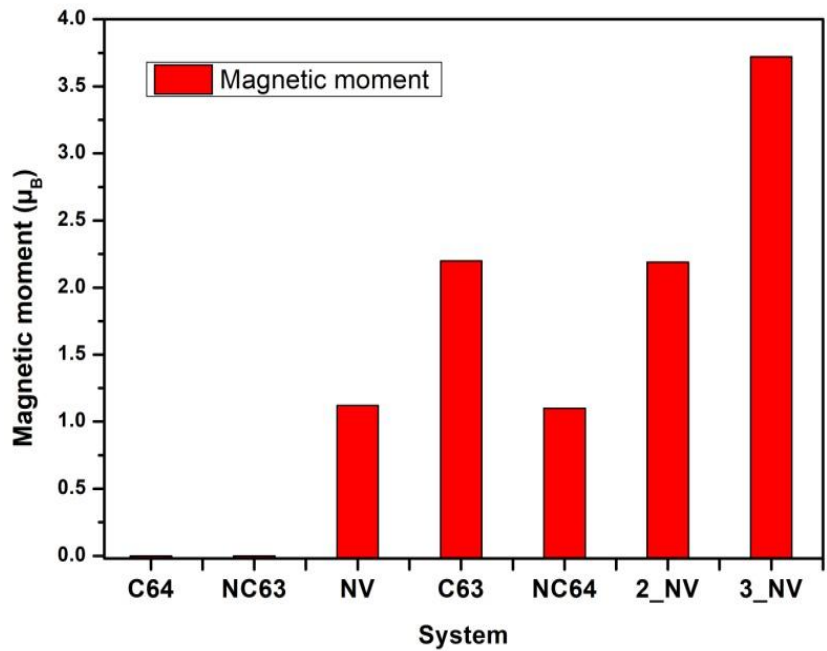

Figure 2: Comparison of the induced magnetic moment in various diamond systems under consideration.

The above result contrasts with earlier reports claiming magnetism in nitrogen diamond. However, this explains the anomaly and occurrence of unconventional magnetism in both nitrogen irradiated nanocrystalline samples [8] and nitrogen doped single crystal diamonds [9]. The present simulation predicts that the generation of a single C-vacancy can generate a magnetic moment of 2.20 $\mu_B$ in the system. The magnetism associated with atomic vacancies has been observed in various samples [26, 27]. Particularly in the diamond system, the generation of C-vacancies creates $sp^2$ carbon around the vacancy, which gives rise to magnetic properties [8]. Additionally, the atomic vacancy generates dangling bonds, which change the electronic and magnetic properties [28]. Similarly, the incorporation of nitrogen at the interstitial site can generate a moment of 1.1 $\mu_B$ in the system. The source of the magnetic moment in nitrogen doped in the interstitial system is the N-atom itself, which is different from the nitrogen present at the lattice position of the NV center. The spin polarised total density of states was plotted and compared for different configuration as shown in Figure 3.

It is interesting to note that spin polarised density of states is symmetric for both the pristine system and nitrogen doped diamond thus generating no net magnetic moment. However, an asymmetric density of state distributions is observed for both carbon vacancies and NV centers, implying an unequal distribution of up spin and down spin states, which gives rise to a net magnetic moment in the system. Moreover, asymmetric half-states (up-states or down-states) are observed around the Fermi energy label. The presence of asymmetric states around the Fermi level makes the system half-metallic [29]. The half metallic character is further enhanced as the concentration of the NV-center is increased as shown in Figure 4. The magnitude of the induced magnetic moment also increased to 2.19 $\mu_B$ and 3.72 $\mu_B$ for 2NV-center and 3NV-center systems, respectively. To identify the source of magnetism, the projected density of states was calculated for the NV center system. The asymmetry in dos comes from the p-orbital electrons of both the carbon and nitrogen atoms surrounding the vacancy. The $p_z$-orbital of the nitrogen doped system has the maximum asymmetry, as shown in Figure 5. Furthermore, the Fermi energy level shifts toward the conduction band in $NC_{64}$, whereas it shifts toward valence band in all other cases, which makes them p-type semiconductors with respect to the pristine system.

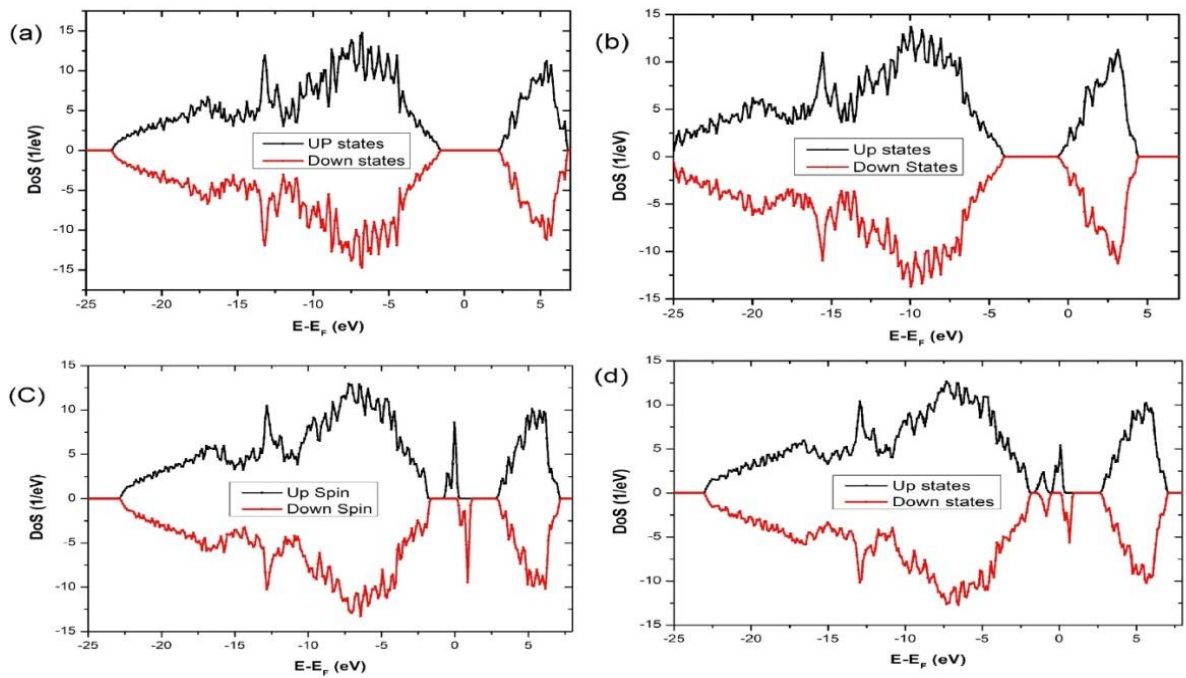

Figure 3: Total density of states in different diamond systems a) pristine, b) nitrogen doped diamond, c) NV-center and d) single C-vacancy.

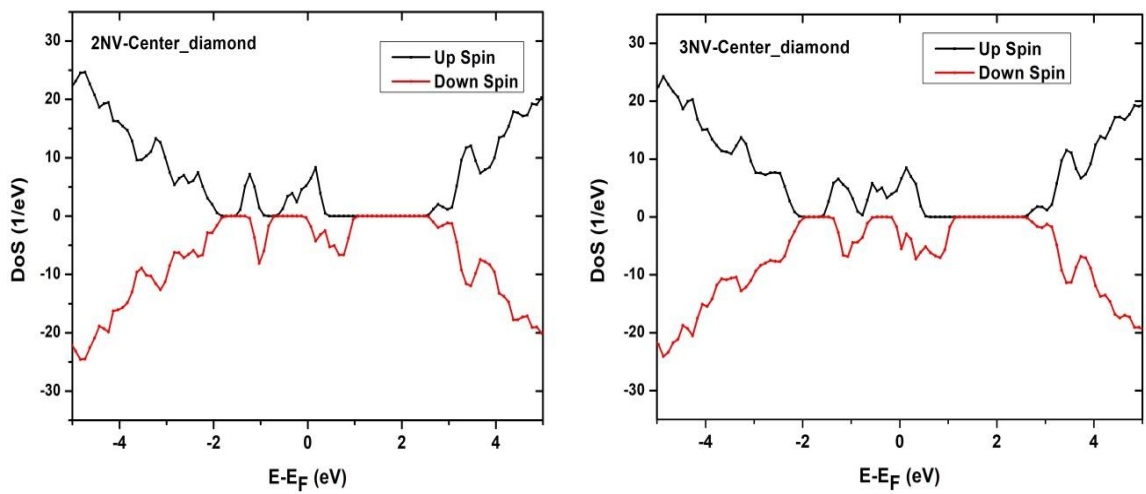

Figure 4: Spin polarised total density of states for the 2NV-center ($N_2C_{140}$) and 3NV-center ($N_3C_{138}$) diamond systems. The half-metallic characteristic is observed around the Fermi level, which has been set to zero.

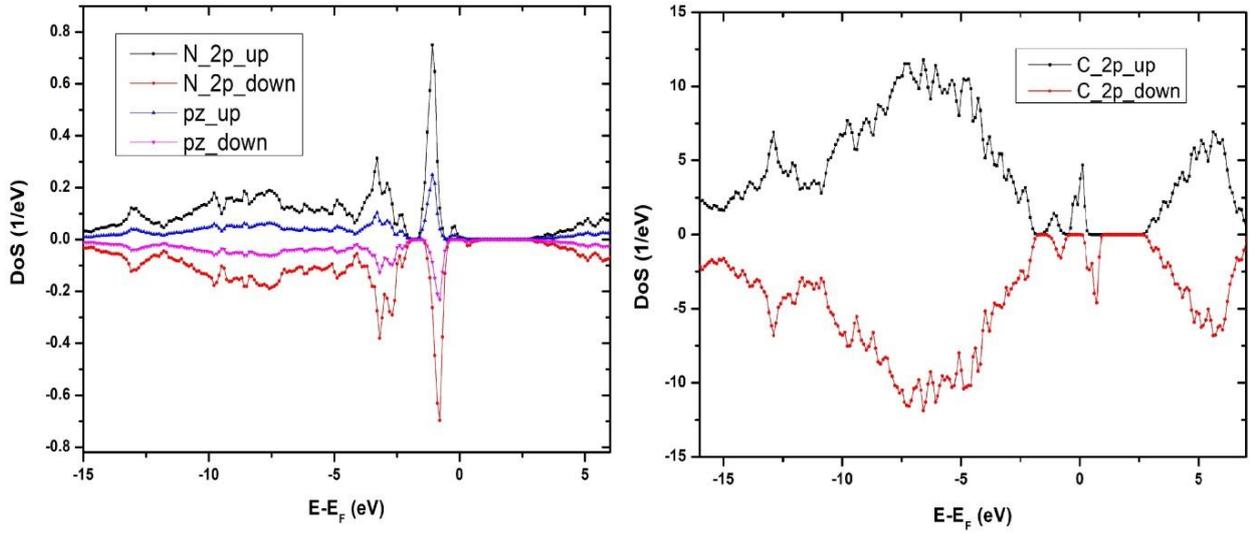

Figure 5: Projected density of states for the 2p-orbitals of the N and C atoms present in the NV center diamond system. The half metallic character is prominent in both the N and C atoms present in the vicinity of C-vacancy in NV-center configuration.

Table 1: Details of the magnetic moment, Fermi energy, Fermi energy shift and defect formation energy in various diamond systems under consideration

| System | Magnetic Moment ($\mu_B$) | Fermi Energy (eV) | Fermi shift w.r.t. Pristine (eV) | Defect Formation Energy (eV) |
|---|---|---|---|---|
| Pristine ($C_{64}$) | 0.0 | 15.07 | -- | NA |
| Nitrogen doped ($NC_{63}$) | 0.0 | 17.44 | 2.37 | 2.83 |
| NV-Centre ($NC_{62}$) | 1.12 | 14.42 | - 0.65 | 5.51 |
| C-vacancy ($C_{63}$) | 2.20 | 14.33 | - 0.74 | 10.59 |
| N-interstitial ($NC_{64}$) | 1.11 | 14.67 | - 0.4 | 2.41 |
| 2NV-Centre ($N_2C_{140}$) | 2.19 | 14.53 | - 0.54 | NA |
| 3NV-Centre ($N_3C_{138}$) | 3.72 | 14.37 | - 0.7 | NA |

During the experimental process, ion implantation by the process of energetic ions irradiation can generate a large numbers of atomic vacancies [13, 27, 30]. Addutionally, nanocrystalline samples many atomic defects at the grain boundaries which can induce magnetic moment [26, 27]. The doping of nitrogen atoms at the interstitial site is energetically more favourable than that at the lattice site, which is in agreement with the findings for other materials [30]. This could also be the origin of magnetism along with C-atom vacancies in earlier experimental reports, where magnetism has been observed in nitrogen doped samples. Moreover, magnetism in nitrogen doped systems has been reported to vanish at temperatures above 25 K [9], which can be understood from the analogy of defect formation energy in each system. The value of the defect formation energy for each system under consideration is tabulated in Table 1. The nitrogen atom present in the interstitial sites has the lowest defect formation energy in diamond, which makes it most viable in spontaneous processes, followed by N-atom placed at the carbon lattice site. The formation energy is maximum for a carbon vacancy which is close to 10.6 eV. The defect formation energy for NV-center system was found to be 5.5 eV, this value is in agreement with earlier report [31]. As the temperature of the system increases, the carbon atomic vacancy may be occupied by interstitial nitrogen atoms, which results in a nonmagnetic $NC_{63}$ configuration. This, explains why the magnetism in the N-doped system was reported to vanish at higher temperatures. The defect formation energy ($E_{DFE}$) for various systems has been calculated via the equation below [13, 32].

$$E_{DFE} = E_D - E_P - \mu_N + \mu_C + q[E_V + E_F + \Delta V]$$

Where $E_D$ is the ground-state total Free energy for the diamond system with defects and $E_P$ is the ground-state total Free energy for the pristine diamond. $\mu_C$ and $\mu_N$ are the chemical potentials of carbon atom and nitrogen atom respectively, $E_F$ is the Fermi energy, $E_V$ is the valence-band maximum in the bulk for undoped system, and $\Delta V$ is the correction term used to align the reference potential in the defect-rich supercell with that in bulk. To obtain the defect formation in each defect bearing system, the number of carbon atoms and nitrogen atoms are changed in the original pristine system which is a 64 carbon atom system.

**Conclusion** – The anomaly of the origin of magnetic properties in nitrogen doped diamond systems has been reinvestigated within the framework of density functional theory. The magnetic moment in the nitrogen doped diamond system actually arises from the presence of carbon vacancies rather than the doping of nitrogen at the lattice site. It is interesting to note that the NV-center also gives rise to a magnetic moment in the system along with the N-atom present at the interstitial site. The magnitude of magnetic moment increases with increasing number of NV-centere in the diamond system. The p-orbital electrons of both C and N atoms present in the vicinity of C-vacancy contribute to the induction of magnetism in NV-center diamond system. The theoretical explanation of the origin of magnetic moment in nitrogen doped diamond has been addressed for the first time. The absence of magnetic properties at higher temperatures in earlier reports has also been correlated with the energy of defect formation in the diamond system. The half-metallic behaviour along with the p-type semiconducting property of NV center diamond will find potential applications in future spintronics devices.

***

*Acknowledgement:* Homnath Luitel dually acknowledges the School of Physics and the University of the Witwatersrand, Johaneshburg for research fellowship under the URC scheme. HL gratefully acknowledges Dr. Robert Warmbier, Professor Somnath Bhattacharyya and Professor Alex Quandt for allowing the use of computer cluster at Wits University and other helpful discussions.

*Data Availability Statement:* The data presented in this work will be made available upon request from the corresponding author.

*Conflicts of Interest*: The authors declare no conflict of interest.